\documentclass{rstransa}

\usepackage{graphicx}
\providecommand{\includegraphics}[2][]{}
\usepackage{amsmath}
\usepackage{amssymb}
\usepackage{mathtools}
\usepackage{hyperref}
\usepackage{cancel}
\usepackage{amsthm}
\usepackage{epigraph}
\newcommand{\Mathematica}{\textit{Mathematica\textsuperscript{\resizebox{!}{0.8ex}{\textregistered}}}}
\def\8{\infty}

\def\const{\textit{ const }}
\def\undertext#1{\vtop{\hbox{#1}\kern 1pt \hrule}}

\def\pd#1{\partial_{#1}}
\def\VEV#1{\left\langle #1\right\rangle}

\def\dbyd#1#2{\frac{d#1}{d#2}}

\def\pbyp#1#2{\frac{\partial#1}{\partial#2}}
\def\ff#1{\frac{\delta}{\delta#1}}

\def\bea{\begin{eqnarray} && &&}
\def\eea{\end{eqnarray}}

\let\oldexp\exp
\renewcommand{\exp}[1]{\oldexp\left(#1\right)}

\def\NS{Navier-Stokes}

\newcommand{\Mod}[1]{\ (\mathrm{mod}\ #1)}

\def\XXint#1#2#3{{\setbox0=\hbox{$#1{#2#3}{\int}$}
     \vcenter{\hbox{$#2#3$}}\kern-.5\wd0}}

\newcommand{\tpmod}[1]{{\@displayfalse\Mod{#1}}}
\usepackage[english]{babel}

\newcommand{\pctPDF}[2]{
\begin{figure}
    \centering
    \includegraphics[width=\textwidth]{#1.pdf}
    \caption{#2}
    \label{fig::#1}
\end{figure}
}

\def\Sc{\mathit{Sc}}
\newcommand{\vect}[1]{\boldsymbol{#1}}
\def\NS{Navier-Stokes}

\def\VEV#1{\left\langle #1 \right\rangle}
\def\I{\mathrm{i}}

\def\pd#1{\partial_{#1}}
\def\ff#1{\frac{\delta}{\delta #1}}

\DeclarePairedDelimiter{\floor}{\lfloor}{\rfloor}

\newtheorem{remark}{Remark}[section]

\makeatletter
\openaccfalse
\makeatother

\begin{document}

\title{Geometric Solution of Turbulent Mixing}

\author{Alexander Migdal}

\address{School of Mathematics, Institute for Advanced Study, 1 Einstein Dr., Princeton, NJ 08540, USA}

\subject{Fluid dynamics, Statistical physics, Theoretical physics} 


\corres{Alexander Migdal\\\email{amigdal@ias.edu}}

\begin{abstract}

We derive an analytic solution for the density of a passive scalar in decaying homogeneous turbulence, in the limit of high Reynolds number and fixed Schmidt number. The velocity statistics are described by the Euler ensemble, previously obtained as a spontaneously stochastic solution of the loop equation associated with the Navier–Stokes equations. The scalar advection–diffusion problem is formulated as a closed linear loop equation and solved within this framework.

For a localized initial condition, the solution consists of a sequence of expanding concentric shells. The radial scalar profile is piecewise parabolic and supported at discrete radii, with amplitudes determined by Euler totients. This structure differs from conventional scaling descriptions of scalar turbulence. Finite diffusivity or sustained forcing smooths the discontinuities while preserving the leading-order geometry.

The results provide a geometric description of scalar transport in decaying turbulence and may be relevant in regimes where dissipation is weak, such as astrophysical or quantum fluids. The predicted shell structure is difficult to resolve directly in numerical simulations; however, its statistical signature is captured by the volume-averaged scalar density, which provides a practical observable.

\end{abstract}

\maketitle
\setlength{\epigraphwidth}{0.7\textwidth}

\epigraph{``To the flows observed in the long run... there correspond certain solutions... These solutions constitute a certain manifold $M$ invariant under phase flow.''}{--- \textup{Eberhard Hopf (1948)}}

\epigraph{``The turbulence of number theory is the statistics of the values of the Euler function $\varphi(n)$... This is a very intermittent sequence.''}{--- \textup{V.I. Arnol'd (1993)}}
\section{Introduction}
\label{preface}
Turbulence—the complex, chaotic motion of fluids—remains one of the central open problems in classical physics. Despite extensive theoretical work and direct numerical simulations (DNS), a predictive and analytically tractable description is still lacking.

This paper presents an analytic solution for passive-scalar mixing in decaying turbulence. It builds on a recent solution for decaying turbulence derived within the strongly turbulent regime using loop calculus. In that formulation, the Navier–Stokes dynamics are recast in loop space, leading to a dual description in which the turbulent state is represented by the Euler ensemble—a spontaneously stochastic probability measure supported on solutions of the Navier–Stokes loop equation. In this limit, discrete structures emerge in momentum-loop space, with a regulator $N \to \infty$ at fixed effective viscosity $\tilde{\nu}=\nu N^2$.

Three conceptual elements underlie this approach:
\begin{enumerate}
    \item \emph{Loop-space formulation.} Instead of the velocity field $\mathbf{v}(\mathbf{r},t)$, we consider circulation along closed loops. This leads to a closed linear equation for the loop functional, analogous to a Schrödinger-type (complex diffusion) equation in loop space.

    \item \emph{Spontaneous stochasticity.} The turbulent regime is described by a statistical solution supported on an attractor in loop space. In the present construction, this attractor has a discrete structure in momentum space, arising in the limit $\nu \to 0$ with $\tilde{\nu}$ fixed.

    \item \emph{Static and liquid loop equations.} The Eulerian (static) loop equation contains both advection and diffusion terms. In the Lagrangian (“liquid”) formulation, the advection term cancels by Kelvin’s theorem. The Euler ensemble provides a solution consistent with both formulations.

    \item \emph{Arithmetic structure.} The resulting statistics exhibit an organization governed by Euler totients and related number-theoretic quantities, which reappear in the scalar solution.
\end{enumerate}

\begin{remark}{\textbf{Euler totients and rational distributions}}
The appearance of the Euler totient function $\varphi(q)$ is related to the distribution of irreducible fractions $p/q$ with bounded denominators. For each $q$, the number of such fractions is $\varphi(q)$, and their total number up to $N$ scales as $\sum_{q=1}^N \varphi(q) \sim \frac{3}{\pi^2}N^2$.

As $N \to \infty$, these fractions become uniformly distributed while retaining nontrivial fine-scale correlations governed by number-theoretic structure \cite{boca2000distribution, marklof2013finescale}. This discrete organization underlies the structures that appear in the loop formulation.
\end{remark}

\begin{remark}{\textbf{Spontaneous stochasticity}}
In conventional turbulence theory, spontaneous stochasticity is often associated with the breakdown of deterministic trajectories due to sensitivity to initial conditions. In the present framework, it refers to the emergence of an invariant statistical measure supported on a turbulent attractor, in the sense of Hopf.

This attractor may be continuous or discrete. In the Euler ensemble constructed here, it has a discrete structure in momentum loop space, and the dynamics correspond to evolution within this invariant measure.
\end{remark}

DNS results \cite{SreeniAkash2025} support several predictions of the Euler ensemble in decaying turbulence. In particular, the decay of turbulent kinetic energy $E(t)\sim t^{-5/4}$ and the scaling $L(t)\sim t^{1/2}$ are consistent with numerical and experimental observations, as are refined statistics such as the scale-dependent behavior of velocity increments.

We extend this framework to passive scalar transport in decaying turbulence. Scalar mixing (e.g. temperature or concentration) is central in many applications, including environmental and astrophysical flows. Conventional approaches often rely on surrogate velocity statistics or closure assumptions, and analytical control becomes limited at high Reynolds number.

The passive scalar evolves according to the advection–diffusion equation in a prescribed velocity field. The problem is to compute scalar statistics averaged over the turbulent ensemble. Previous analytical approaches frequently employ Gaussian surrogate models, such as the Kraichnan model \cite{PassiveScalarKraicknan}, which allow exact calculations but do not fully capture the dynamics of the Navier–Stokes system.

In contrast, the loop formulation leads to a closed linear equation for the scalar loop functional, which can be solved directly at fixed Schmidt number in the decaying regime.

\begin{remark}{\textbf{Closure problem and surrogate models}}
The advection term couples scalar correlations to velocity statistics, generating an infinite hierarchy of moment equations. Analytical treatments typically rely either on closure assumptions or on simplified velocity models.

In the present approach, the loop formulation provides a closed linear equation for the scalar functional without introducing surrogate statistics or truncating moment hierarchies.
\end{remark}

Within the Euler ensemble, a localized initial condition produces a hierarchy of expanding concentric shells. The radial scalar profile is piecewise parabolic and supported at discrete radii, with amplitudes determined by Euler totients. Finite diffusivity or sustained forcing smooths this structure.

Although the resulting discontinuities are difficult to resolve directly, their statistical signature is captured by volume-averaged observables, providing a practical target for DNS verification.

\subsection*{Paper outline}

Section~\ref{turbmix} introduces the governing equations and the decaying regime. Section~\ref{liquidloop} summarizes the loop formulation and the Euler ensemble for the velocity field. Section~\ref{scalarEq} derives the scalar loop equation and presents its solution. Further discussion is given in \cite{ReviewPaperAM}. Conclusions are presented in Section~\ref{conclusion}.

\section{The turbulent mixing problem}\label{turbmix}

Turbulent mixing—the transport and diffusion of a passive scalar by a turbulent velocity field—is a central process in fluid dynamics, with applications ranging from environmental flows to astrophysical systems \cite{sreenivasan2018turbulent,dimotakis2005turbulent,warhaft2000passive}.

The governing equations are the incompressible Navier--Stokes equations coupled to the scalar advection--diffusion equation:
\begin{subequations}\label{basicEqs}
\begin{align}
&\frac{\partial \vect{v}}{\partial t} + (\vect{v} \cdot \nabla)\vect{v} = -\nabla p + \nu \nabla^2 \vect{v}, \\
&\nabla \cdot \vect{v} = 0, \\
&\frac{\partial T}{\partial t} + (\vect{v} \cdot \nabla) T = D \nabla^2 T,
\end{align}
\end{subequations}
where $\vect{v}(\vect{r}, t)$ is the velocity field, $p$ the pressure, $T(\vect{r}, t)$ the scalar field, $\nu$ the kinematic viscosity, and $D$ the scalar diffusivity. The dimensionless Schmidt number is $\mathrm{Sc}=\nu/D$.

Passive-scalar turbulence has been extensively studied using global observables such as variance decay, spectra, and correlation functions (see, for example, \cite{PassiveScalarKraicknan}). A central question concerns the spatial structure generated by a localized initial condition in decaying turbulence. Conventional descriptions typically predict Gaussian spreading under the combined action of advection and diffusion, but do not provide a closed-form solution in this regime.

In this work, we consider scalar transport in a velocity field drawn from the Euler ensemble—a spontaneously stochastic probability measure supported on solutions of the unforced Navier--Stokes equations in the vanishing-viscosity limit \cite{migdal2024quantum}. Within this framework, the advection--diffusion problem admits a closed linear formulation in loop space, which can be solved without introducing surrogate velocity statistics or closure assumptions.

For a localized initial condition, the resulting scalar field develops a sequence of concentric spherical shells. The radial profile is piecewise parabolic and supported at discrete radii, with amplitudes determined by Euler totients. Finite diffusivity smooths these discontinuities while preserving the leading-order structure. This behaviour arises as a property of the Euler ensemble in the decaying regime and may be relevant in systems where dissipative effects are weak.

\section{Decaying turbulence and the Euler Ensemble}\label{liquidloop}

The velocity statistics underlying the passive scalar problem are taken from the exact solution of decaying turbulence obtained via loop calculus \cite{migdal2023exact, migdal2024quantum, migdal2025duality, ReviewPaperAM}. In this formulation, the Navier–Stokes dynamics are expressed in terms of loop observables, leading to a linear equation in loop space. In the vanishing-viscosity limit, the corresponding attractor is the Euler ensemble \cite{ReviewPaperAM}. We summarize the elements required for the scalar problem.

\subsection{Static vs.\ liquid loops and the turbulent attractor}

Loop dynamics may be formulated either in an Eulerian frame (fixed loop $C$) or in a Lagrangian frame (loop advected with the flow). The circulation
$$
\Gamma_C[v]= \oint_C d \vect r \cdot \vect v(\vect r, t)
$$
evolves differently in the two cases.

In the Eulerian frame:
\begin{align}\label{staticEq}
    &\partial_t \Gamma =  \oint_C d \vect r \cdot \left(-\nu\vect \nabla \times \vect \omega + \vect v \times \vect \omega \right),\\
    & \omega = \vect \nabla \times v;
\end{align}

In the Lagrangian frame, Kelvin’s theorem removes the advection term:
\begin{align}
& \partial_t \vect C(\theta,t) = \vect v\left(\vect C(\theta,t),t\right),\\
& \partial_t \Gamma_C[v] = -\nu \oint_C d \vect r \cdot \vect \nabla \times \vect \omega.
\end{align}

Both formulations admit the same attractor in the decaying, vanishing-viscosity limit \cite{migdal2023exact, ReviewPaperAM}. This attractor, the Euler ensemble $\mathcal{E}$, is a statistical measure supported on a discrete set of momentum-loop configurations. The solution satisfies the liquid loop equation by construction, and the advection term in the static equation vanishes upon integration for this ensemble. This allows one to work in the liquid-loop formulation without loss of generality.

\subsection{Loop calculus framework and the Euler ensemble}

The loop-space formulation employs functional derivatives with respect to the loop velocity $\mathbf{C}'(\theta)$ to represent differential operators acting on loop functionals \cite{MMEq79, Mig83, Mig98Hidden, migdal2025YangMills, ReviewPaperAM}. The loop functional
$$
\Psi[C,t] = \left\langle \exp{\I \Gamma_C[v] / \nu} \right\rangle_v
$$
satisfies a linear evolution equation.

In the liquid formulation:
\begin{equation}\label{LoopEq}
\partial_t \Psi[C, t] = \nu \mathcal{L}_C \Psi[C, t],
\end{equation}
with
\begin{align}\label{diffusionOperator}
   & \mathcal L_C = \oint d \theta C'_\nu(\theta) \hat L_\nu(\theta),\\
   &\hat L_\nu(\theta) = T^{\alpha\beta\gamma}_\nu \frac{\delta^3}{\delta C'_\alpha(\theta-0)\delta C'_\beta(\theta)\delta C'_\gamma( \theta+0)},\\
   &T^{\alpha\beta\gamma}_\nu =\delta_{\alpha\beta}\delta_{\gamma\nu}+ \delta_{\gamma\beta}\delta_{\alpha\nu}-2\delta_{\alpha\gamma}\delta_{\beta\nu}.
\end{align}

The ordering of functional derivatives reflects the presence of vorticity. The resulting discontinuities are kinematical and arise even for smooth velocity fields.

\subsection{Momentum loop equation}

A functional Fourier transform leads to momentum loop variables:
\begin{align}\label{Anzatz}
    & \Psi[C,t] = \left\langle \exp{\I \int_0^{2\pi}  d \theta \vect C'_\mu(\theta) \cdot \vect P(\theta,t)} \right\rangle_{P(t)},\\
\label{MLE}
    & \frac{1}{\nu}\partial_t\vect P(\theta) = (\Delta \vect P \cdot \vect P) \Delta \vect P-(\Delta \vect P)^2 \vect P,\\
    &  \vect P(\theta) =  \frac{\vect P(\theta+)+\vect P(\theta-)}{2}, \quad
    \Delta \vect P = \vect  P(\theta+)- \vect P(\theta-).
\end{align}

The discontinuity $\Delta \vect P$ encodes vorticity. Momentum loops must therefore be treated as distributions rather than continuous functions.

\begin{remark}{\textbf{Statistical averaging}}
The averaging over $P(\theta,t)$ may be interpreted either as an initial statistical distribution or as averaging over a turbulent attractor in the sense of Hopf. In the present framework, the attractor is identified with the Euler ensemble, which provides a discrete statistical measure in momentum-loop space.
\end{remark}

\subsection{Euler ensemble solution}

In the limit $\nu \to 0$ with $\tilde{\nu} = \nu N^2$ fixed, the solution takes the form
\begin{align}
    &\vect P(\theta) = \frac{\vect F(\theta)}{\sqrt{2\nu\,(t+t_0)}},\\
    \label{Feq}
    &  (\Delta \vect F^2-1)\vect{F} = (\Delta \vect F  \cdot \vect F )  \Delta \vect F .
\end{align}

\begin{remark}{\textbf{Time translation}}
The parameter $t_0$ reflects time-translation invariance and sets the origin of time in comparison with experiments or simulations.
\end{remark}

The solutions form a manifold $\mathcal{E}$ corresponding to random walks on regular star polygons. The discrete parameters of these walks are distributed according to number-theoretic constraints.

\subsection*{The Euler ensemble as a turbulent attractor}

In this construction, $\vect F(\theta)$ arises as the continuum limit of a random walk on a regular polygon. The geometry enforces $|\Delta \vect F|=1$ and $\vect F \cdot \Delta \vect F=0$, which solves \eqref{Feq}.

The ensemble is defined by averaging over all such configurations. The closure condition requires rational winding angles, leading to a counting governed by Euler totients. This provides the statistical structure of the ensemble.

\subsection{Connection to observables}

The loop formalism allows computation of physical observables through functional derivatives of $\Psi[C,t]$. The Euler ensemble yields explicit predictions for quantities such as energy decay $E(t)\sim t^{-5/4}$ and integral scale growth $L(t)\sim t^{1/2}$ \cite{ReviewPaperAM}. These agree with experimental and numerical results in the decaying regime.

In what follows, this Euler ensemble is taken as the statistical description of the velocity field and used to derive the passive scalar solution.

\section{Passive scalar dynamics in loop space}\label{scalarEq}

In the conventional approach \cite{PassiveScalarKraicknan}, the advection--diffusion equation is coupled to an unknown velocity field; progress typically comes from Gaussian surrogates or ad hoc closures of the moment hierarchy. In the loop formulation, the problem reduces to a closed linear equation for the loop functional, which we solve exactly—no surrogates, no closures.

\subsection{circulation and temperature joint distribution}

Consider the passive scalar (temperature) evaluated at a loop point $\vect r_1=\vect C(\theta_1)$. The advection--diffusion equation can be written in terms of loop operators as
\begin{align}\label{Teq}
    \pd{t} T(\vect C(\theta_1),t) = \left(-\vect v(\vect C(\theta_1)) \cdot \pd{\vect C(\theta_1)} +D \left(\pd{\vect C(\theta_1)}\right)^2 \right)T(\vect C(\theta_1),t)
\end{align}
Note that the gradient $\pd{\vect C(\theta_1)}$ represents the singular limit in which only a single point on the loop is shifted
\begin{equation}
    \pd{\vect C(\theta)} = \ff{\vect C'(\theta-0)} - \ff{\vect C'(\theta+0)};
\end{equation}
The last formula for the gradient was proven in loop space calculus (Appendix A of \cite{ReviewPaperAM}) using a conventional functional derivative
\begin{align}
   & \pd{\vect C(\theta)} = \int_{\theta-0}^{\theta+0} d \theta' \ff{\vect C(\theta')} = -\int_{\theta-0}^{\theta+0} d \theta' \pd{\theta'}\ff{\vect C'(\theta')} =\ff{\vect C'(\theta-0)} - \ff{\vect C'(\theta+0)};
\end{align}
In momentum loop space, this discontinuity of the functional derivative translates to a discontinuity of the momentum loop.

Because the scalar does not feed back on the flow, the Euler ensemble continues to describe the velocity statistics. Diffusion affects only the scalar statistics, so we must solve the advection–diffusion problem in loop space given by \eqref{Teq}, and then average the resulting solution over the distribution of the velocity field that follows from the turbulent solution of the \NS{} equation \eqref{basicEqs}.
This task seems impossible within the direct approach to these equations, but with the loop space approach, these obstacles can be bypassed.

Let us consider the mixed average
\begin{align}\label{TaverageNS}
    &T_C(\theta_1,t) = \VEV{T\left(\vect C(\theta_1), t\right) \exp{\frac{\I \Gamma_{C}[v]}{\nu} }}_{v}
\end{align}
which correlates the temperature at $\vect C(\theta_1)$ with the circulation around $C$. Unlike the mean temperature $\VEV{T(\vec r,t)}_{v}$, which is a function of a single point in space (plus time), this $T_C(\theta_1,t)$ is a functional of the loop $C$.  This step is the first in our ascent to the loop space from the physical $ 3+1$-dimensional space.

Differentiating in time produces contributions from both \eqref{Teq} and the \NS{} dynamics:
\begin{align}\label{TCderiv}
    &\pd{t} T_C(\theta_1,t) = \VEV{\exp{\frac{\I \Gamma_{C}[v]}{\nu} }\left(-\vect v(\vect C(\theta_1)) \cdot \pd{\vect C(\theta_1)} +D \pd{\vect C(\theta_1)}^2 \right)T\left(\vect C(\theta_1), t\right) }_{v}\nonumber\\
    &+ \nu \VEV{T\left(\vect C(\theta_1), t\right) \mathcal L_C \exp{\frac{\I \Gamma_{C}[v]}{\nu} }}_{v}
\end{align}
We are going to take advantage of the fact established in Appendix A of \cite{ReviewPaperAM} that the advection term cancels in the Euler ensemble solution for the \NS{} loop equation, so we were left only with the diffusion term in loop operator $\mathcal L_C $, \eqref{LoopEq}.
\subsection{The loop equation for the passive scalar}

The vorticity and velocity fields are represented in loop space as operators
\begin{align}
& \hat \omega(\theta) = -\I\nu \ff{\vect \sigma(\theta)} = -\I\nu \ff{\vect C'(\theta-)}\times \ff{\vect C'(\theta+)};\\
 &   \hat v(\theta_1) = \frac{-1}{\pd{\vect C(\theta_1)}^2} \pd{\vect C(\theta_1)} \times \hat \omega(\theta);
\end{align}
In order to rewrite \eqref{TCderiv} as a loop equation for $T_C$, we need to extend the operators on all the factors inside averaging $\VEV{}_v$ over \NS{} dynamics.
Taking the loop operators outside the average $\VEV{\cdot}_v$ would, in principle, generate cross terms when $\pd{\vect C(\theta_1)}$ acts on $\exp{\frac{\I \Gamma_{C}[v]}{\nu} }$ and when $\mathcal L_C $ acts on $T(\vect C(\theta_1),t)$. All such cross terms vanish by the loop calculus identities proven in Appendix A, sections (h) and  (i)  of \cite{ReviewPaperAM}. 
Thus, we get a linear diffusion-advection equation in loop space
\begin{align}\label{TGammaEq}
      &\pd{t} T_C(\theta_1,t) = \left(-\vect v(\vect C(\theta_1)) \cdot \pd{\vect C(\theta_1)} +D \pd{\vect C(\theta_1)}^2 + \nu \mathcal L_C \right)T_C(\theta_1,t)\
\end{align}
\begin{remark}{\textbf{The origin of cancellations.}}

The extra terms all cancel due to the Leibniz property of the loop operator $\mathcal{L}$. Although it involves three functional derivatives, $\mathcal{L}$ effectively acts as a first-order differential operator. This behavior arises because it is constructed from the antisymmetric tensor part of the second functional derivative (corresponding to the vorticity operator $\hat{\omega}$) evaluated at a discontinuity in the loop parameterization (corresponding to the gradient operator $\partial$). When this singular operator is applied to an exponential or a product of functionals, the usual mixed second-derivative terms---arising from cross products of first derivatives---identically vanish. They are annihilated either by the contraction with the antisymmetric tensor or by the absence of the requisite time discontinuity. The rigorous technical details provided in \cite{ReviewPaperAM} essentially formalize this geometric intuition.
\end{remark}
\subsection{Solution of the passive scalar loop equation}

The solution for the above equation has the form of an average over the Euler ensemble
\begin{align}\label{PLoopT}
    T_C(\theta_1,t) = \VEV{\tau(\vect C(\theta_1),t)\exp{\frac{\I \oint d \theta \vect C'(\theta) \cdot \vect F(\theta)}{\sqrt{2 \nu (t+ t_0) }}}}_{\mathcal E}
\end{align}
where $\tau$ is determined by
\begin{align}
    &\pbyp{\tau(\vect C(\theta_1),t)}{t}  \exp{\frac{\I \oint d \theta \vect C'(\theta) \cdot \vect F(\theta)}{\sqrt{2 \nu (t+ t_0) }}} =\nonumber\\
    & \left(-\hat v(\theta_1) \cdot \pd{\vect C(\theta_1)} +D \pd{\vect C(\theta_1)}^2\right) \left(\tau(\vect C(\theta_1),t)\exp{\frac{\I \oint d \theta \vect C'(\theta) \cdot \vect F(\theta)}{\sqrt{2 \nu (t+ t_0) }}} \right);
\end{align}
and the remaining term $\nu \mathcal L_C$ in \eqref{TGammaEq} cancel by the momentum loop equation \eqref{MLE} for $\vect P = \vect F/\sqrt{2 \nu(t+ t_0)}$.

With extra terms coming from applying the operator $\pd{\vect C(\theta_1)}$ to the exponential, we obtain a linear evolution for $\tau$,
\begin{align}\label{taueq}
    &\pbyp{\tau(\vect C(\theta_1),t)}{t} = \hat Q(\theta_1, t)\tau(\vect C(\theta_1),t);\\
    & \hat Q(\theta, t) = \left(\frac{-2  \I D \Delta \vect F(\theta)}{\sqrt{2 \nu(t+ t_0)}}-\tilde v(\theta,t)\right) \cdot \pd{\vect C(\theta)} +D \pd{\vect C(\theta)}^2 - D \frac{(\Delta \vect F(\theta))^2}{2 \nu (t + t_0)} ;\\
    & \tilde v(\theta,t) = \left.\hat v(\theta_1,t)\right|_{\mathcal E}
\end{align}
Here $\left.\hat v(\theta_1,t)\right|_{\mathcal E}$ is the velocity operator in momentum loop space, with momentum loop according to the Euler ensemble
\begin{align}
    \left.\hat v(\theta_1)\right|_{\mathcal E} =  \frac{1}{\Delta \vect F^2(\theta_1)} \Delta \vect F(\theta_1) \times ( \vect F(\theta_1) \times \Delta \vect F(\theta_1)) \frac {\nu}{\sqrt{2\nu(t + t_0)}} = \frac {\nu \vect F(\theta_1)}{\sqrt{2\nu(t + t_0)}};
\end{align}
\subsection{The discretization of the solution}
The strict mathematical meaning of these discontinuities can be provided by replacing the loop $C$ with a polygon, corresponding to the equidistant angles $\theta_k = 2 \pi k/N$ on a unit circle. This discretization was used in \cite{DeLellisInprep} to justify the loop calculus and prove that the Euler ensemble solves it at finite $N$ (theorem 10.6 in their paper). This regulator must tend to infinity in final observables, as is done in quantum field theory. 

Then, with $\vect F_k \equiv \vect F(\theta_k) $
\begin{align}
     & \Delta \vect F(\theta_k)  =  \vect F_k - \vect F_{k-1};
\end{align}
With $|\Delta \vect F(\theta_k)|^2=1$ one finds
\begin{align}
       &\tilde v(\theta_k,t) = \frac {\nu \vect F_{<k>} }{\sqrt{2\nu(t + t_0)}};\\
       & \vect F_{<k>} = \frac{\vect F_k + \vect F_{k-1}}{2};
\end{align}
This simple differential equation is solvable in the Fourier representation
\begin{align}\label{FourierTemp}
 & \tau(\vect r, t) = \int \frac{d^3 q}{(2 \pi)^3} \exp{\I \vect q \cdot \vect r} \eta(\vect q, t);
\end{align}
Substituting this Anzatz into \eqref{taueq} we  get linear ODE for $\eta(\vect q, t)$
\begin{align}\label{etaeq}
    &\dbyd{ \eta(\vect q, t)}{t} = \left(\vect q \cdot \frac{2   D  \Delta \vect F_k - \I \nu \vect F_{<k>}}{\sqrt{2 \nu(t+ t_0)}}  -D \vect q^2 -\frac{D}{2 \nu (t + t_0)}\right)\eta(\vect q, t)
\end{align}  
which we readily solve:
\begin{align}\label{etasol}
    & \eta(\vect q, t) =W(\vect q) \left(\frac{t + t_0}{t_0}\right)^{-\frac{1}{2\Sc}} \exp{- \frac{\vect q^2 \nu t}{\Sc} -\I \left(\sqrt{2 \nu (t + t_0)}-\sqrt{2 \nu t_0}\right)\vect q  \cdot \vect G_k};\\
    & \vect G_k = \frac{\vect F_{k} + \vect F_{k-1}}{2}+ 2\I\frac{\vect F_{k}- \vect F_{k-1}}{\Sc}  
\end{align}
Here $\Sc=\nu/D$ is the Schmidt number, the sole dimensionless parameter for the scalar.

The weight $W(\vect q)$ reflects the initial scalar distribution:
\begin{align}
     W(\vect q) = \eta(\vect q,0) = \int d^3 r \exp{-\I \vect q \cdot \vect r} T(\vect r,0)
\end{align}
The Fourier quantity $\eta(\vect q,t)$ is complex because the temperature is averaged with the phase factor $\exp{\I \Gamma_C[v]/\nu}$.

The mean temperature is, of course, real and positive. To extract it, contract the loop to a point: set $\vect r=\vect C(\theta_1)$ and $\vect C'(\theta)=0$. Then $\Gamma_C[v]=0$, and the original average \eqref{TaverageNS} reduces to the scalar mean at a point. On the right of \eqref{PLoopT}, setting $\vect C'=0$ eliminates the momentum-loop factor. Thus
\begin{align}\label{TbeforeLimit}
    &\VEV{T(\vect r,t)}_{v} = \VEV{\tau(\vect C(\theta_1),t)}_{\mathcal E} = \left(\frac{t + t_0}{t_0}\right)^{-\frac{1}{2\Sc}} \nonumber\\
    &\int \frac{d^3 q}{(2 \pi)^3} W(\vect q)\exp{\I \vect q \cdot \vect r}\VEV{ \exp{- \frac{\vect q^2 \nu t}{\Sc} -\I \left(\sqrt{2 \nu (t + t_0)}-\sqrt{2 \nu t_0}\right)\vect q  \cdot \vect G_k}}_{\mathcal E};
\end{align}

\subsection{The turbulent limit. Harmony of primes behind chaos}
As shown in \cite{migdal2023exact,migdal2024quantum}, the turbulent limit combines infinitesimal viscosity with large polygon radius:
\begin{align}
   & \nu \to \frac{\tilde \nu}{N^2} = O(1/N^2);\\
   &  R \to \rho N  = O(N);\\
   & \vect F(\theta) \to N \rho \left\{\cos \alpha, \sin \alpha,0\right\};\\
   & |\Delta \vect F(\theta)| = 1;
\end{align}

\begin{remark}{\textbf{The physical meaning of the turbulent viscosity $\tilde{\nu}$.}}

The reader may question the origin of the residual finite viscosity scale $\tilde{\nu}$, interpreting it as an odd phenomenological artifact. On the contrary, it is the rigorous mathematical manifestation of dynamical renormalization in the turbulent limit.

The formal turbulent limit of the \NS{} equations is often mathematically idealized as vanishing viscosity ($\nu \to 0$). In physical reality, of course, the kinematic viscosity $\nu$ is a fixed, measurable fluid parameter with the dimensions of velocity circulation ($L^2/T$). The Reynolds number (or, more precisely, the Reynolds functional) represents the ratio of the macroscopic velocity circulation to this microscopic molecular viscosity. When studying the highly turbulent regime, we are examining the phenomenon where this ratio is extremely large. Physically, this is achieved by pumping more energy into the fluid to scale up the vortical part of the velocity and provoke macroscopic vortices, rather than by artificially scaling down the physical viscosity, which remains a fixed material parameter.

Within our loop space formulation, a continuous turbulent loop is mathematically approximated by a polygon with $N$ sides. The scale $\sqrt{\nu t}$ represents the typical microscopic viscous diffusion length, which is vanishingly small compared to the typical macroscopic vortex size in the turbulent limit. However, the fluid dynamics along the loop are cumulative. By multiplying this microscopic diffusion length by the number of polygon sides $N$, we dynamically generate an aggregated macroscopic length scale, $N\sqrt{\nu t}$, that is directly comparable to the typical loop sizes in our solution. 

We require this macroscopic combination to remain strictly finite in the continuum limit $N \to \infty$. This leads directly to nontrivial correlation functions that depend naturally on the scaling variable $r/\sqrt{\tilde{\nu} t}$, governed by the dynamically renormalized turbulent viscosity $\tilde{\nu} = N^2 \nu$. Therefore, rather than being an \emph{ad hoc} phenomenological parameter, $\tilde{\nu}$ reveals exactly how a vanishingly small bare viscosity becomes macroscopically renormalized by the infinite geometric complexity of the turbulent loops to produce finite, observable dissipative effects in the Euler ensemble solution.
\end{remark}
In this limit, the vector $\sqrt{\nu}\vect G_k$ tends to a finite real vector,
\begin{align}
    & \sqrt{\nu}\vect G_k \to \sqrt{\tilde \nu} \rho \hat \Omega \cdot\left\{\cos \alpha, \sin \alpha,0\right\};
\end{align}
The distribution of $\rho=\big(2N\sin(\pi p/q)\big)^{-1}$ for coprime $p,q$ with $0 < p < q < N \to \infty$ was found in \cite{migdal2024quantum} by some advanced methods of number theory. This is a discontinuous, piecewise power-like distribution:
 \begin{align}\label{rhodist}
    & f(\rho)= \left(1- \frac{\pi ^2}{21600 \zeta (5)}\right) \delta(\rho) + \frac{2\pi^3}{3} \rho^4\Phi\left(\floor*{\frac{1}{2\pi \rho}}\right);
\end{align}
with $\Phi$ the totient summatory function:
\begin{align}
    \label{PhiDef}
    &\Phi(q) = \sum_{n=1}^q \varphi(n);\\
     & \varphi(m) = m \prod_{p|m}\left(1 - \frac{1}{p}\right)
\end{align}

For a point-source initial condition,
\begin{align}
     T(\vect r,0) = T_0 \delta(\vect r)
\end{align}
we find from \eqref{FourierTemp}
\begin{align}
    &\VEV{T(\vect r, t)}_v = T_0(1+t/t_0)^{-\frac{1}{2\Sc}} G(r,t);\\
    & G(r,t) = \VEV{\delta\left(\vect r -\rho \left(\sqrt{2 \tilde \nu (t+ t_0)} -\sqrt{2 \tilde \nu  t_0}\right) \hat \Omega \cdot\left\{\cos \alpha_k, \sin \alpha_k,0\right\}\right)}_{\rho, \hat \Omega} 
\end{align}
The one-dimensional delta in \eqref{rhodist} produces a three-dimensional delta in $G$; the remaining term yields a finite structure. Averaging  $G(r,t)$  over $\hat\Omega \in\mathbb O(3)$ removes the dependence upon $\alpha_k$ leading to a simple answer
\begin{align}\label{Grt}
 &     G(r,t) = \left(1- \frac{\pi ^2}{21600 \zeta (5)}\right)\delta(\vect r) + \frac{\pi^2}{6  r^2} \int d \rho \rho^4 \delta\left(r- 2\pi\rho L(t)\right)\Phi\left(\floor*{\frac{1}{2\pi \rho}}\right);\\
  & L(t) = \frac{\sqrt{2 \tilde \nu (t+ t_0)} -\sqrt{2 \tilde \nu  t_0}}{2 \pi};
\end{align}
and after integrating over $\rho$,
\begin{align}
    &\VEV{T(\vect r, t)}_v =  (1+t/t_0)^{-\frac{1}{2\Sc}}\left[\left(1- \frac{\pi ^2}{21600 \zeta (5)}\right)\delta(\vect r) + \frac{r^2 \Phi\left(\floor*{\frac{L(t)}{r}}\right)}{768 \pi^5 L(t)^5}\right];
\end{align}
This is a decaying prefactor multiplying a rescaled delta function plus a second, decaying contribution with a \emph{discontinuous} dependence on the scaling variable $r/L(t)$ (Fig.~\ref{fig::PhiXiPlot}). The support lies at $r\le L(t)$, so the shells contract towards the origin as $t\downarrow 0$.

Thus, a point source produces {figureinfinitely many concentric spherical shells, $\frac{L(t)}{n}<r<\frac{L(t)}{n-1}$, accumulating at the origin. The profile is parabolic within each shell, with jumps $\propto \varphi(n)/n^2$ at successive boundaries. Since these jumps decrease with $n$, the temperature approaches a finite limit at the origin:
\begin{align}
     \lim_{n\to \8}\frac{\Phi\left(n\right)}{n^2} = \frac{3}{\pi^2}
\end{align}
The discontinuities reflect the absence of a continuum limit for Euler’s totient; e.g.,
\begin{align}
   &\varphi (7919) = 7918;\\
   &\varphi (7920) = 1920;
\end{align}
although statistically, for large $n$,
\begin{align}
     \varphi(n) \sim \const{} \frac{n}{\log \log n} \quad \text{for almost all } n
\end{align}

Observing such a shell structure directly in DNS would require very large grids; in physical flows, it may be smoothed by weak dissipation or forcing. Predicted structure of discrete shells with sharp interfaces might explain the 'ramp-cliff' patterns observed experimentally in passive scalar fields in turbulent flows\cite{sreenivasan2018turbulent} (see also Fig.~5 in \cite{ReviewPaperAM}), which feature similar asymmetric profiles.

\pctPDF{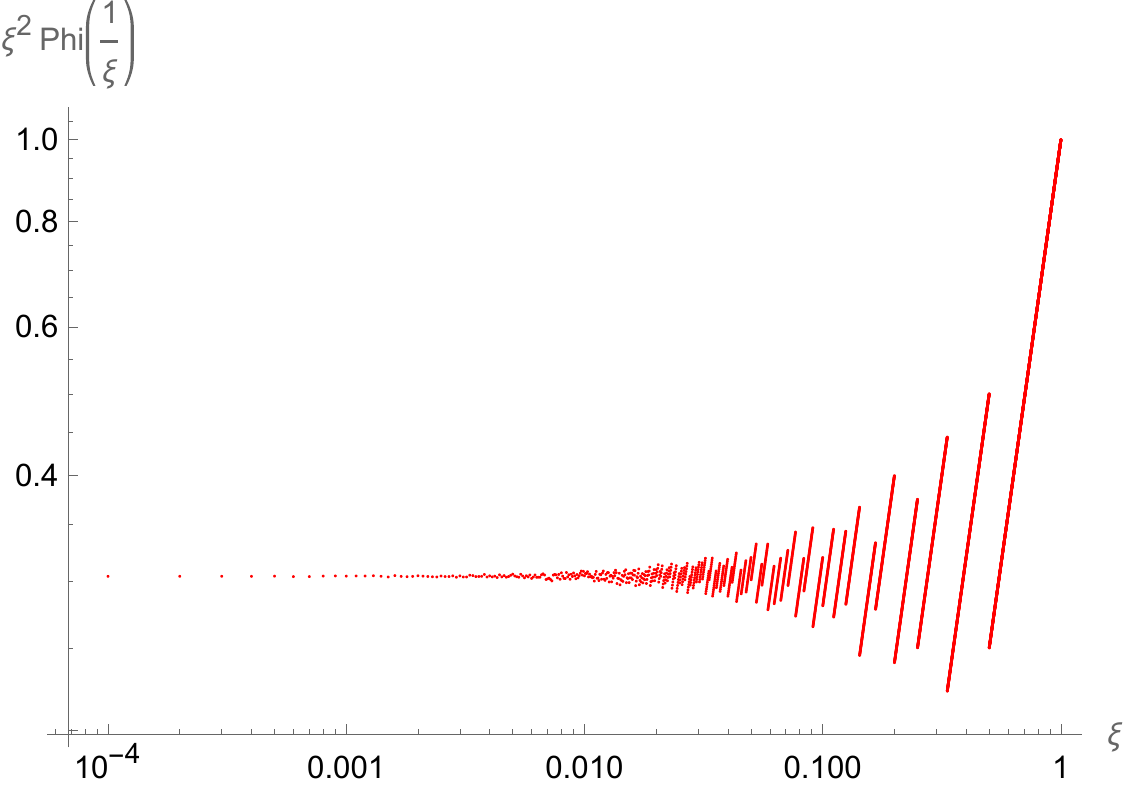}{Log--log plot of the universal function $\xi^2\Phi\left(\floor*{\frac{1}{ \xi}}\right)$ where $\xi = \frac{2 \pi r}{\left(\sqrt{2 \tilde \nu (t+ t_0)} -\sqrt{2 \tilde \nu  t_0}\right)}$.}

For spectral properties, take $W(\vect q')=\delta(\vect q-\vect q')$ (a single Fourier mode):
\begin{align}
    &\VEV{\tilde T(\vect q, t)}_v =T_0\left(1+\frac{t}{t_0}\right)^{-\frac{1}{2\Sc}} U\left(|\vect q| \left(\sqrt{2\tilde \nu (t+ t_0)}-\sqrt{2\tilde \nu t_0}\right)\right);\\
    & U(\kappa) = \VEV{\exp{\I \rho \vect \kappa \cdot \hat \Omega \cdot \left\{\cos \alpha_k, \sin \alpha_k,0\right\}}}_{\rho, \hat \Omega} = \VEV{\frac{\sin(\rho \kappa)}{\rho\kappa}}_{\rho}
\end{align}
Using \eqref{rhodist} and the Taylor expansion of $\sin(\rho\kappa)/(\rho\kappa)$ yields an absolutely convergent series
\begin{align}\label{UkappaExp}
     U(\kappa) =1+ \sum_{l=1}^\8 \frac{(l+1) (2 \pi )^{-2 l-3} \left(-\kappa ^2\right)^l \zeta (2 (l+2))}{3 (2 l+5) \zeta (2 l+5) \Gamma (2 l+3)}
\end{align}
We have evaluated this series numerically in \Mathematica\ with high precision: $U(\kappa)$ starts at $1$, decreases slightly, and then oscillates about $0.99986$ (Fig.~\ref{fig::UKappaPlot}).
\pctPDF{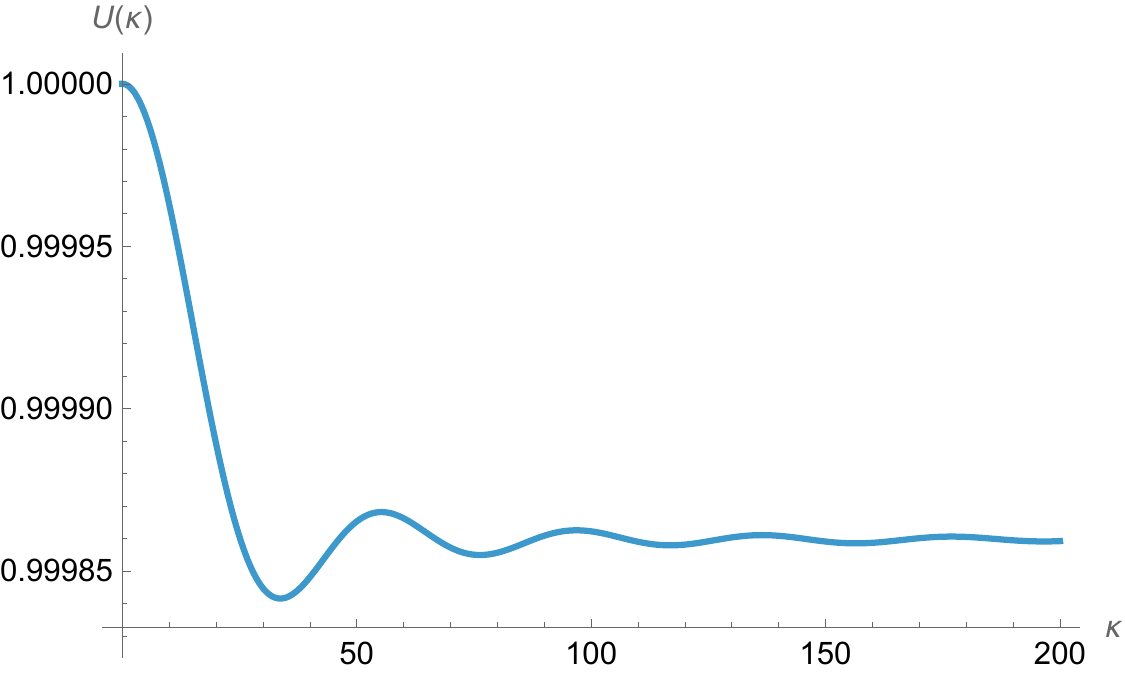}{The plot of universal function $U(\kappa)$ in \eqref{UkappaExp} describing the amplitude of passive scalar plane wave with wavevector $\vect q$ at the time $t$ as a function of scaling variable $\kappa = |\vect q| \left(\sqrt{2\tilde \nu (t+ t_0)}-\sqrt{2\tilde \nu t_0}\right)$}
\subsection{Solution mechanism in loop space}

This approach provides an analytic solution to the passive scalar problem, which is difficult to address directly. A direct formulation would require averaging the scalar field $T(\mathbf{r}, t)$, which depends on the velocity $\mathbf{v}$, over the unknown statistical distribution of $\mathbf{v}$ itself.

We circumvented the direct problem by moving to \textbf{loop space}. Instead of focusing on the velocity field directly, we leverage the \emph{known diffusion laws} governing the evolution of arbitrary liquid loops advected by the flow. We consider the scalar value $T$ at a specific point on such a loop, passively moving with it.

The crucial step is to define a \textbf{joint loop functional} that incorporates both the scalar value at a point and the circulation around the entire loop,  $\VEV{T(\vect C(\theta_1), t) \exp{\frac{\I \Gamma_{C}[v]}{\nu} }}_{v}$. When the governing equations for both the scalar (advection-diffusion) and the velocity (Navier-Stokes, expressed via circulation dynamics) are transformed into loop space operators acting on this joint functional, a remarkable simplification occurs: the resulting evolution equation becomes \textbf{linear}.

Within this linear functional equation, the statistical properties of the velocity field, as captured by the Euler ensemble solution $\Psi[C,t]$ for the circulation functional, effectively become part of the known linear operator. Solving this equation, via functional Fourier transforms (leading to momentum loop space), yields the evolution of the joint functional. The desired average scalar value $\VEV{T(\mathbf{r}, t)}_v$ is then recovered by evaluating this solution in the limit of a contracted loop (equivalent to setting $\vect C'(\theta)=0$). Note that in this limit, the phase factor $\exp{\I \frac{\Gamma_C}{\nu}} $ tends to $1$, and so does the factor $\exp{\I \oint d \theta \vect C'(\theta) \cdot \vect P(\theta,t)} $. The role of these factors was to participate in time differentiation and loop differentiation while the loop was still finite. In the end, the factors disappeared, but their contribution to the loop equation remained. This juggling of the loop, which after differentiation shrinks to zero circulation, is the standard technique used in loop calculus to obtain ordinary correlation functions \cite{migdal2024quantum}.

Essentially, the knowledge of the velocity distribution and the functional dependence of $T$ on $\mathbf{v}$ is implicitly encoded within the loop diffusion laws derived from the fundamental equations. By ascending from the three-dimensional physical space to the infinite-dimensional loop space, the inherent nonlinearity of the original advection problem is avoided, rendering it solvable with linear methods, albeit requiring less familiar mathematical techniques. In this way, the loop-space formulation avoids the direct nonlinearity of the original problem and renders it tractable, in a manner analogous to the linearization of integrable systems via Lax representations and inverse scattering methods \cite{Lax1968, Gardner1967, Zakharov1972}.

\subsection*{Why discrete shells survive at fixed Sc}
We now explain, in standard scaling terms, why discrete shells appear and persist at fixed $\Sc$.
In the decaying, high–Re limit we take $\nu\!\to\!0$ and $D\!\to\!0$ at fixed
$\mathrm{Sc}=\nu/D$. Kelvin cancellation leaves a closed loop equation whose
nonlinear balance sustains finite vorticity and a finite advective speed (of order
unity), whereas the diffusive speed scales as $v_D\sim\sqrt{D/t}\ll 1$. Thus, advection
dominates mixing, with diffusion acting as a secondary smoothener.

In this regime the scalar is transported ballistically to radii $r=\rho\,L(t)$, with
$L(t)\sim \sqrt{t}$ and a dimensionless transport ratio $\rho$ that does not vanish
as $\nu\!\to\!0$. The admissible values of $\rho$ form a discrete set fixed by a
closure/periodicity constraint; for each denominator $q$, the number of admissible
numerators (coprime to $q$) is $\varphi(q)$ (Euler’s totient). Because the
advection–diffusion equation is linear in $T$, the ensemble superposes these discrete
transports, producing concentric shells at $r=L(t)/n$ with parabolic profiles inside
each shell and jumps proportional to $\varphi(n)/n^2$ across shells. Any finite $D$
rounds the discontinuities but leaves the discrete radii and the jump law intact at
leading order. In Fourier variables, this structure appears as a universal function of
$|\mathbf q|\,\big(\sqrt{t+t_0}-\sqrt{t_0}\big)$ with small, parameter–free
oscillations.

A constructive realization of the admissible transport ratios is given in the momentum–loop formulation (see \cite{ReviewPaperAM}), but it is not needed for the scaling argument above.
\medskip 

For a broader discussion relating these loop-space findings to classical turbulence paradigms, including Kolmogorov scaling, multifractal models, and the energy cascade narrative in the context of decaying turbulence, the reader is referred to Section 8 of our companion review \cite{ReviewPaperAM}.
\subsection*{The challenges of observing the spherical shells in DNS}
The density distribution for a source, which starts as a delta function at the initial moment, is impossible to observe in DNS because the singularities cannot be resolved on the lattice. Even ignoring the delta function (which remains with some weaker weight at later stages of advection-diffusion), the lack of isotropy presents a statistical problem. Unlike correlation functions, which only depend on the difference between points in homogeneous isotropic turbulence, this density is not uniform, but rather tied to a specific point. To obtain statistics, one should run a parallel simulation in which the initial point is placed at random positions, and then average the results in relative coordinates. The practical alternative is to compute DNS statistics for the Fourier coefficients of the temperature. This transformation will provide effective averaging over the whole grid in coordinate space, adding statistics. 

Besides, the predicted time dependence of the temperature amplitude in Fourier space is smooth, which would make it easier to compare with DNS data. There is a universal function depending on a scaling variable $\kappa =|\vect q| \sqrt{2\tilde \nu}(\sqrt{t+ t_0} - \sqrt{t_0})$ . Therefore, by fitting $\tilde \nu, t_0$ from the velocity spectrum, where the same parameters appear, one can pool all the statistics in the turbulent range of $|\vect q|$ for each time  $t$ (the range found by studying the velocity spectrum). After this pooling, the resulting scaling function can be matched with our $U(\kappa)$ in \eqref{UkappaExp}.

\subsection*{Volume average of passive scalar}
\pctPDF{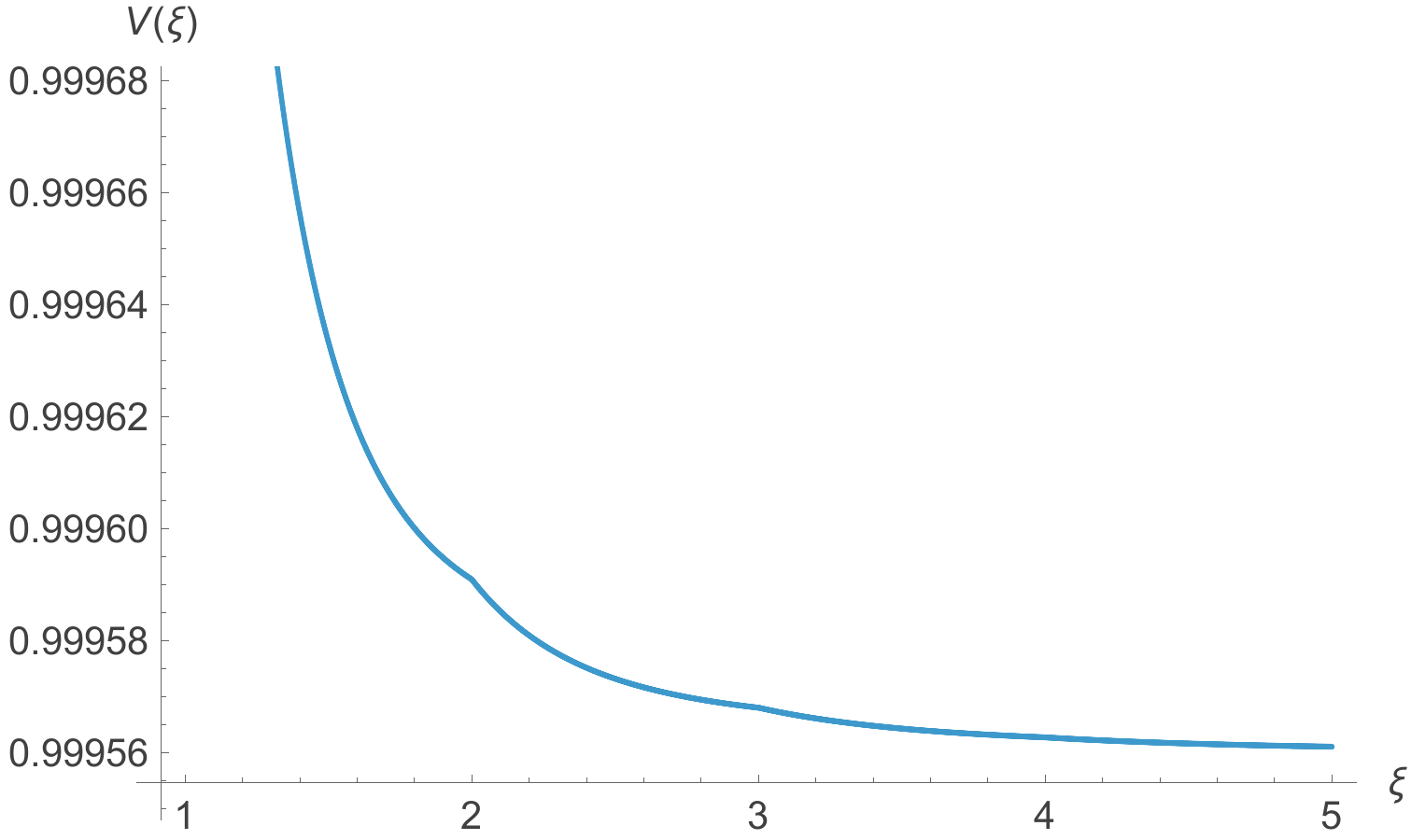}{The plot of universal function $V(\xi)$ in \eqref{GlobalAv} describing the average temperature as a function of  the time $t$ as a function of scaling variable $\xi = \left(\sqrt{2\tilde \nu (t+ t_0)}-\sqrt{2\tilde \nu t_0}\right)/r_0$, where $r_0$ is the radius of the spherical averaging domain. Note the slight discontinuity of the derivative at $\xi=2, 3,...$, when there is one more term added to the sum. This universal function is almost constant, because of the small coefficient $\frac{1}{240 \pi ^2} $.}
Another way to improve statistics in DNS for passive scalars is to study volume averages.
Let us consider the average temperature inside a spherical domain of radius $r_0$ centered at the initial point-like source.
We have to take our point-like solution \eqref{Grt} and compute its mean value inside a spherical ball
\begin{align}
  & \bar G(r_0,t) = \frac{3}{4 \pi r_0^3}\int d^3\vect r \Theta(r_0 -|\vect r|)G(r,t) =\nonumber\\
  & \frac{3}{4 \pi r_0^3}\left(1- \frac{\pi ^2}{21600 \zeta (5)}\right) + \frac{\pi^2}{2  r_0^3} \int_0^{\frac{r_0}{2 \pi L}} d \rho \rho^4 \Phi\left(\floor*{\frac{1}{2\pi \rho}}\right);
\end{align}
Using the definition of the Euler summatory function \eqref{PhiDef}, we obtain the following series
\begin{align}
    &\bar G(r0,t) = \frac{3}{4 \pi r_0^3}\left(1- \frac{\pi ^2}{21600 \zeta (5)}\right) + \frac{\pi^2}{2  r_0^3} \sum_{k=1}^\infty\varphi(k) \int_0^R d \rho \rho^4;\\
    & R = \min{ \left(\frac{1}{2 \pi k}, \frac{r_0}{2\pi L}\right)}
\end{align}
The integral terms can be arranged as follows
\begin{align}
    &\frac{\pi^2}{2 r_0^3} \sum_{k=1}^\infty\varphi(k) \int_0^R d \rho \rho^4 =
    \frac{\pi^2}{10  r_0^3} \Theta(L-r_0)\sum_{k=1}^{\floor{\frac{L}{r_0}}} \varphi(k)\left( \left(\frac{r_0}{2\pi L}\right)^5-(2 \pi k)^{-5} \right) +\nonumber\\
   & \frac{\pi^2}{10  r_0^3} \sum_{k=1}^\infty \varphi(k) (2 \pi k)^{-5} 
\end{align}
The last sum reduces to the zeta function
\begin{align}
    \frac{\pi }{28800 r_0^3 \zeta (5)}
\end{align}
and cancels the similar term in $\bar G(r_0,t)$. Collecting all terms, we find
\begin{align}\label{GlobalAv}
    &  \bar T(r_0,t) = \frac{T_0}{V_0}\left(1+\frac{t}{t_0}\right)^{-\frac{1}{2\Sc}}V\left(\frac{\sqrt{2\tilde \nu (t+ t_0)}-\sqrt{2\tilde \nu t_0}}{r_0}\right);\\
    & V_0 = \frac{4 \pi r_0^3}{3};\\
    & V\left(\xi\right)=1 + \frac{\Theta(\xi-1)}{240 \pi^2} \sum_{k=1}^{\floor{\xi}} \varphi(k)\left( \xi^{-5}-k^{-5} \right) 
\end{align}
This universal function $V(\xi)$ is still not mathematically smooth because of the floor value taken as an upper limit of the sum, but it is much smoother than the original shell structure.  (see Figure \ref{fig::PassiveScalarVPlot}). It is numerically very close to one, which would not be hard to verify in DNS.
\begin{remark}{\textbf{MHD as an example of active advection/diffusion.}}

The reader may wonder whether these methods can incorporate matter backreaction (such as gravity). Indeed, the loop space framework is fully applicable to ``active'' fields that exert dynamical feedback on the fluid. 

A prime example is Magnetohydrodynamics (MHD), where the magnetic field reacts back onto the velocity field via the Lorentz force. As demonstrated in \cite{Migdal2025MHD}, the exact statistics of MHD turbulence can be solved using the loop equation. This active feedback leads to a dual formulation consisting of two interacting Euler ensembles—one governing the hydrodynamic circulation and the other the magnetic circulation. Remarkably, this exact geometric solution dynamically generates a phase transition precisely at the magnetic Prandtl number $\mathrm{Pr}_m = 1$. This confirms that the loop calculus is not inherently restricted to passive scalars but can rigorously accommodate complex, nonlinear feedback mechanisms. This example indicates that the loop calculus extends beyond passive scalars and can accommodate systems with dynamical feedback.

\end{remark}
\subsection*{Implications of the Shell Structure for the Millennium Problem}

The analytical solution for the passive scalar, detailed in Section 4(b) and illustrated in the figures above, also raises questions relevant to the Navier–Stokes regularity problem. This framework, which describes the evolution of a scalar field coupled to the Navier-Stokes equations, presents an explicit, analytical solution characterized by a finite-time singularity in the coupled NS-advection system.

This singularity is of a highly unusual nature. It is not a singular point, line, or surface, but rather a \textbf{fractal set of concentric, spherical discontinuities}. It is notable that Tao also conjectured a fractal singularity, albeit via a different mathematical mechanism \cite{Tao2016, Klarreich2014Fluid}. The scalar density profile is not smooth; it consists of piecewise-parabolic profiles separated by finite, discontinuous jumps. In the limit, this intermittent structure condenses into a delta function.

Crucially, the geometric and statistical properties of this singular structure are dictated by number theory. The radial locations and jump magnitudes of the shells are \textbf{governed by the Euler totient summatory function}, rooted in the distribution of prime numbers. This provides a physical realization of the deep mathematical intuition described by V.I. Arnol'd: ``The turbulence of number theory is the statistics of the values of the Euler function $\varphi(n)$... This is a very intermittent sequence...'' \cite{Arnold1993Statistics}. While Arnol'd proposed this as a mathematical analogy, the Euler ensemble solution provides a dynamical mechanism deriving this exact structure from the fundamental equations of fluid dynamics.

This solution relates \textit{statistical} intermittency to spatial singularities through number-theoretic structures (Farey sequences \cite{franel1924farey}). The spectrum of decay exponents obtained in the Euler ensemble includes complex values of the form $7 \pm \I t_n$, where $1/2 \pm \I t_n$ are the nontrivial zeros of the Riemann zeta function. These exponents enter directly into the time and space dependence of velocity correlation functions derived in \cite{migdal2024quantum}. The imaginary parts generate oscillatory corrections to the decay laws, which translate into measurable modulations in spectral observables. Such oscillatory behaviour has been reported in recent high-Reynolds-number experiments and DNS \cite{ReviewPaperAM}, and is consistent with this structure.

\paragraph{Disclaimer: Physics vs.\ Formal Regularity}

We do not claim a solution to the Millennium Prize Problem as formulated by the Clay Mathematics Institute. That problem concerns the existence and smoothness of individual deterministic solutions of the Navier–Stokes equations for all time, given smooth initial data.

The present work is formulated within the framework of theoretical physics, where the relevant notion of a “solution” to a chaotic dynamical system is its invariant measure. In the case of turbulence, this corresponds to the statistical ensemble (the Euler ensemble), rather than a single deterministic realization.

From a physical perspective, the assumption of perfectly smooth initial data is an idealization. In any physical fluid at nonzero temperature, thermal fluctuations imply that the initial state is described by a statistical ensemble rather than a single smooth configuration. Consequently, the notion of global regularity for an individual realization is not directly tied to physically observable behavior, particularly in the high Reynolds number regime.

Our work, therefore, addresses the statistical physics of turbulent flows. The loop calculus provides a description of the invariant measure governing turbulence, in which individual realizations are subordinate to ensemble properties. This is complementary to, but distinct from, the mathematical question of global regularity posed in the Millennium Prize Problem.

\section{Conclusions}\label{conclusion}

We have obtained an analytic solution for the one-point scalar distribution from a point source in decaying turbulence, in the joint limit $\nu \to 0$, $D \to 0$ at fixed Schmidt number $\Sc = \nu/D$. The background velocity statistics are described by the Euler ensemble. In this regime, scalar transport is dominated by advection, while molecular diffusion provides subleading corrections. The mean field develops compact, approximately spherical shells with outer radius scaling as $L(t) \sim \sqrt{t}$.

\medskip
\noindent\textbf{Main implications.}

\begin{itemize}

\item \textit{Closed scalar loop equation and analytic solution.}  
The loop formulation yields a closed linear equation for the scalar loop functional and admits an analytic solution at fixed $\Sc$. In real space, a point source generates a hierarchy of concentric shells at discrete radii $r = L(t)/n$, with piecewise parabolic profiles. The discontinuities at the shells scale as $\varphi(n)/n^2$, reflecting number-theoretic multiplicities. Finite diffusivity smooths these discontinuities but preserves the discrete radii at leading order.

\item \textit{Equivalence of liquid and static loop formulations.}  
The Euler ensemble satisfies the liquid (Lagrangian) loop equation. On this solution, the advection term in the static (Eulerian) loop equation vanishes, so both formulations are consistent. A scaling argument at fixed $\Sc$ shows that the advective contribution remains of order unity, while diffusion is suppressed, explaining the persistence of the shell structure.

\item \textit{Fourier-space signature.}  
For monochromatic initial data, the solution reduces to a universal response function $U(\kappa)$ with an absolutely convergent expansion in $\kappa = |\mathbf{q}|\sqrt{2\tilde{\nu}}(\sqrt{t+t_0}-\sqrt{t_0})$. The amplitude decays as $(1+t/t_0)^{-1/(2\Sc)}$. The function $U(\kappa)$ exhibits small oscillations that provide a practical observable in DNS, without requiring resolution of real-space discontinuities.

\item \textit{Consistency with decaying velocity statistics.}  
The scalar solution is based on the Euler ensemble, which reproduces known decay laws: $E(t)\sim t^{-5/4}$ and $L(t)\sim t^{1/2}$, together with a continuously varying scaling exponent for $\langle (\Delta v)^2 \rangle$. No adjustable dimensionless parameters are introduced.

\item \textit{Verification strategy.}  
Direct observation of shells may be difficult at finite resolution. However, the Fourier-space observable $U(\kappa)$ and the volume-averaged scalar profile $V(\xi)$ provide robust quantities for numerical verification.

\item \textit{Range of validity.}  
The analysis applies to unforced, decaying turbulence in the high-Reynolds-number limit with vanishing diffusivity at fixed $\Sc$. Finite diffusivity or weak forcing smooths the discontinuities, while the discrete structure persists in the pre-asymptotic regime. Questions of uniqueness and stability of the statistical state remain open.

\item \textit{Early-time behaviour.}  
For $t \ll t_0$, the decay factor $(1+t/t_0)^{-1/(2\Sc)}$ is well approximated by $\exp{-t/(2t_0\Sc)}$. Long-time evolution is therefore required to separate transient exponential behaviour from the asymptotic algebraic regime.

\end{itemize}

\medskip
\noindent\textbf{Outlook.}

Several extensions follow naturally. These include general initial conditions through the weight $W(\mathbf{q})$, two-point scalar statistics and fluxes, dependence on $\Sc$ and weak inhomogeneity, and the stability of the shell structure under forcing. On the mathematical side, a more detailed analysis of the diffusion contribution and its relation to the discrete structure would be of interest.

On the numerical side, targeted DNS measurements of $U(\kappa)$, and where possible the shell radii and jump amplitudes, would provide direct tests of the predictions. It is also of interest to explore the relation between the predicted shell structure and the experimentally observed ramp--cliff features in turbulent scalar fields \cite{sreenivasan2018turbulent}.

\section{ACKNOWLEDGEMENTS}
We gratefully acknowledge insightful discussions with Semon Rezchikov, Edoardo Spezzano, Katepalli Sreenivasan, and James Stone.

This research was partly supported by the Simons Foundation (award ID SFI-MPS-T-MPS-00010544) at the Institute for Advanced Study.
\section*{Declaration of Interests}
The author reports no conflict of interest.
\section*{Declaration of AI use}
I used AI-assisted writing tools (Gemini, GPT-5) solely for stylistic editing and text formatting of this manuscript. The tools were employed to improve readability and flow, standardize terminology and section headings, and convert punctuation to LaTeX-safe forms. No scientific content, derivations, equations, symbols, figures, data analysis, or results were generated or altered by the tool. I manually reviewed and verified all edited text for accuracy and consistency with the submitted manuscript. Any errors remain my responsibility.

\bibliographystyle{plainnat}
\bibliography{bibliography}

\end{document}